\documentclass[iop,apj]{emulateapj}

\usepackage{xspace}
\usepackage{hyperref}
\newcommand{\vmax}{$V_{\rm max}$\xspace}

\shorttitle{Constraing Dwarf Quenching}
\shortauthors{Slater \& Bell}

\slugcomment{}

\begin{document}

\title{Confronting Models of Dwarf Galaxy Quenching \\
with Observations of the Local Group}

\author{Colin T. Slater and Eric F. Bell}
\affil{Department of Astronomy, University of Michigan,
    500 Church St., Ann Arbor, MI 48109; \href{mailto:ctslater@umich.edu}{ctslater@umich.edu}}

\begin{abstract}
A number of mechanisms have been proposed to connect star-forming dwarf
irregular galaxies with the formation of non-star-forming dwarf spheroidal galaxies,
but distinguishing between these mechanisms has been difficult. We use the Via
Lactea dark matter only cosmological simulations to test two well-motivated
simple hypotheses---transformation of irregulars into dwarf spheroidal galaxies
by tidal stirring and ram pressure stripping following a close passage to the
host galaxy, and transformation via mergers between dwarfs---and predict the
radial distribution and inferred formation times of the resulting dwarf
spheroidal galaxies. We compare this to the observed distribution in the Local
Group and show that 1) the observed dSph distribution far from the Galaxy or M31
can be matched by the VL halos that have passed near the host galaxy at least
once, though significant halo-to-halo scatter exists, 2) models that require two
or more pericenter passages for dSph-formation cannot account for the dSphs
beyond 500 kpc such as Cetus and Tucana, and 3) mergers predict a flat radial
distribution of dSphs and cannot account for the high dSph fraction near the
Galaxy, but are not ruled out at large distances. The models also suggest that
for dSphs found today beyond 500 kpc, mergers tend to occur significantly
earlier than dwarf--host encounters, thus leading to a potentially observable
difference in stellar populations. We argue that tidal interactions are
sufficient to reproduce the observed distribution of dSphs if and only if a
single pericenter passage is sufficient to form a dSph.
\end{abstract}


\section{Introduction}

The origin of the approximate dichotomy between the star-forming dwarf irregular
galaxies (dIrrs) and the non-star-forming, pressure supported dwarf spheroidal
galaxies (dSphs) has long been an open question
\citep{hodge69,faber83,kormendy85,gallagher94}.
Several mechanisms have been proposed to create dSphs, such as tidal stirring
and stripping \citep{mayer01,klimentowski09,kazantzidis11a,kravtsov04}, resonant
stripping \citep{donghia09}, or ram pressure stripping \citep{mayer06}. This
broad grouping of models all involve the influence of a large host galaxy, which
is motivated by the observed trend in the Local Group for most dSphs to be found
within 200-300 kpc of either the Milky Way or M31
\citep{vandenbergh94,grebel03}. Other theories for dSph formation do not require
the influence of a larger galaxy and transform dIrrs into dSphs via either
heating of the dwarfs' cold gas by the UV background \citep{gnedin00}, strong
feedback \citep{dekel86,maclow99,gnedin02,sawala10}, or mergers between dwarfs
at early times \citep{kazantzidis11b}.

Since many of these mechanisms can all be shown to plausibly produce dSphs given
the right initial conditions, it can become difficult to distinguish between
these theories as many leave only weak signatures on the individual galaxies.
Because of this limitation, one might alternatively study the signatures these
processes leave on the population of Local Group dwarfs as a whole. The orbital
and assembly histories of the dwarfs in the Local Group vary significantly, and
the link between these histories and the resulting morphologies of the dwarfs
may be a telling indication of which processes are at work. This perspective
aims for differentiating between mechanisms for dSph formation where they are
most different---where and when they act---rather than where they are all
generally similar in the injection of energy into the orbits of stars in
the dwarf.

In this work we use cosmological simulations \citep[Via Lactea I \& II,][herein
VL1 and VL2]{diemand07,diemand08} to trace the histories of the dwarfs that
survive to today, and use these histories to infer which dwarfs (in aggregate)
may have been affected either by tidal stirring or by mergers between dwarfs.
The large difference in when and where these mechanisms act on dwarfs creates
significant differentiation in the resulting distribution of dSphs. These two
cases are also particularly suitable for study with high resolution
dark-matter-only simulations, since the behavior of the luminous components can
be inferred from the behavior of the dark matter.  That is, we can infer the
effects of tidal forces or mergers experienced by a galaxy by tracking the dark
matter halo and applying relatively simple criteria based only on the halo
properties. These criteria are physically motivated based on controlled
simulations of the individual processes
\citep[e.g.,][]{kazantzidis11b,kazantzidis13}. Clearly these simulations will
predict some detailed properties of the dwarfs that will not be captured by our
binary dSph-or-not criteria, but our focus on the bulk properties of the dwarf
population as a whole will minimize the impact of these differences on our
conclusions. These simplifications enable us to understand the formation of dSph
galaxies in a broader cosmological context rather than only in controlled
experiments. 

Much of this work focuses on the dwarfs currently outside the virial radius of
the Galaxy (or M31). The distribution of distant satellites that were once found
inside the virial radius of a host has been investigated before in simulations
of cluster or group environments \citep{balogh00,moore04,gill05,wetzel13} and
Milky Way-like environments \citep{diemand07,teyssier12}. The existence of such
galaxies is well established. Similarly, the rate and timing of mergers between
dwarfs in a Milky Way-like environment has been studied with simulations
\citep{klimentowski10}, but comparisons to observations have remained limited.
Our work focuses on bringing both of these mechanisms for the formation of dSphs
to a specific comparison with the observed distribution of Local Group dwarfs.

Towards that goal, we discuss the simulations and our criteria for both
interactions and major mergers in Section~\ref{sims}, and present the results
and a comparison to the observed dSph distribution in the Local Group in
Section~\ref{obs}. The distribution of times at which galaxies either merge or
experience close passages is described in Section~\ref{times}, and we discuss
the implications of these results in Section~\ref{conclusions}.

\section{Analysis of Simulations}
\label{sims}

We use both the Via Lactea simulation \citep{diemand07} and Via Lactea
II\footnote{\url{http://www.physik.uzh.ch/~diemand/vl/}} \citep{diemand08} for
our analysis of tidal interactions, and only the VL2 simulation for our analysis
of mergers. Both are cosmological, dark matter only simulations centered on a
Milky Way-sized halo with a virial mass of $1.93\times 10^{12} M_\odot$ in VL2
($1.77 \times 10^{12} M_\odot$ in VL1), corresponding to a virial radius
($r_{200}$) of $402$ kpc ($389$ kpc in VL1). VL1 used $234 \times 10^6$
particles of mass $2 \times 10^4$ $M_\odot$, while VL2 had $1.1 \times 10^9$
particles each of mass $4.1 \times 10^3$ $M_\odot$. Both simulations are
entirely sufficient to resolve all of the luminous observable satellites,
and we will generally restrict our results to halos with a maximum circular
velocity (\vmax) greater than 5 km/s at $z=0$. At this limit halos have an
average of 350 particles in VL1 and 800 particles in VL2. Dark matter halos
were identified in the simulation using a phase-space friends-of-friends (6DFOF)
algorithm, as described in detail in \citet{diemand06}. These halos were then
linked across snapshots by identifying halos which share significant numbers of
particles; in identifying the most massive progenitor at least 50\% of the
particles in the descendant are required to be present in the progenitor, and
conversely 50\% of the progenitor particles to be present in the descendent.
This constraint is later relaxed when computing merger trees, but the process is
similar. Note that 6DFOF only links the central, low energy particles together;
that is, the fraction of common particles between the progenitor and the
descendant is usually significantly larger among the 6DFOF particles than among
all particles within the virial radius.  

In addition to the Milky Way-analog halo (referred to as the ``main'' halo for
convenience), in VL2 there is also a second large galaxy present in the simulations
that happens to have properties similar to Andromeda. This was identified in
\citet{teyssier12}, who refer to it as ``Halo 2'' and showed that it has a total
gravitationally-bound mass of $6.5 \times 10^{11}$ $M_\odot$, and lies $830$ kpc
from the main halo. Both of these properties are conveniently similar to
Andromeda, and as a result, when we discuss the interaction between dwarf
galaxy-sized halos and a massive host, we consider either the main halo or Halo
2 to be sufficient for this purpose. Ignoring Halo 2 would significantly bias
our results, since dwarf galaxy halos that become bound to it may experience
substantial tidal interactions while their distance from the main halo is still
large. Treating both large halos on an equal footing also reflects our treatment
of the observed Local Group dwarfs, where we consider the dwarfs' distance to
either the Milky Way or Andromeda, whichever is less. VL1 has no such analogous
component, so we do not apply the same conditions.

\subsection{Tidal Interactions}

With the evolutionary tracks of halos in place, we can identify halos that are
strong candidates to have undergone some form of interaction with a larger
galaxy. This process is similar to that of \citet{teyssier12} but not identical.
Of the several thousand most massive halos identified at $z=0$, we select only
those with \vmax values between $5$ and $35$ km/s. This cut conservatively
ensures that the halos we track are well resolved, and broadly spans the \vmax
values of classical dwarf galaxies. The position of the halo's most massive
progenitor is then tracked back through each snapshot, and both it and the
position of the two host halos are linearly interpolated between snapshots. The
pericenter distance and the number of pericenter passages between the halo and
either host is then recorded. While interpolation between timesteps is not
ideal, it does provide some assurance that we are not substantially
overestimating the minimum radius of each pericenter passage by only taking the
distance at individual snapshots. We have verified that this interpolation
produces accurate results for VL2 (where the larger timesteps make it more
important) by using the more densely sampled Via Lactea 1 simulation,
downsampling the timesteps to the VL2 resolution and testing the interpolation.
The results show the interpolation works particularly well for the distant halos
we focus on here as most of them are on strongly radial, fly-by trajectories. 

The distance between the halo and either of the hosts is then compared to the
virial radius ($r_{200,{\rm mean}}$, defined to enclose a density 200 times the
cosmic mean density) of the main galaxy as a function of redshift, and the
minimum of this ratio is found. This establishes the depth to which the halo has
reached in a large galaxy. We assume the virial radius of Halo 2 is the same as
that of the main halo, and we later show that our results are not particularly
sensitive to the exact radius criterion. We also track the number of pericentric
passages the halos have undergone inside of $R_{vir}/2$ of the host halo by
finding minima in the halo-host distance. The resulting halo statistics are in
good agreement with those obtained by \citet{teyssier12}; out of all selected
halos, a very large majority (96\%) have at some point been inside of half the
host virial radius, and approximately 11\% of those that have been inside this
radius are later found at $z=0$ outside of virial radius.

We note that our model relies on the assumption that the halos in the
simulations are populated with observable dwarf galaxies in an unbiased way.
This assumption is potentially called into question by the ``missing satellites
problem'', which may suggest that the number of subhalos in simulations is
substantially larger than the number of dwarf galaxies in the Local Group
\citep{moore99,klypin99}. For example, in VL1 and VL2 we include 9992 and 2224
halos, respectively, in our analysis, but only 101 observed dwarfs \citep[from
the catalog of][]{mcconnachie12}. This discrepancy can be plausibly resolved within
the cold dark matter framework by a combination of observational incompleteness
and supression of star formation in small halos, thus decreasing their
luminosity below detectability in current surveys
\citep{somerville02,koposov09}, or by (additionally) destroying or diminishing
the mass of halos through tidal stripping \citep{kravtsov04,brooks13}. Our use
of ratios of number counts of halos limits our sensitivity to models that alter
the mapping between halos and dwarfs based only on their mass, since the motion
of the halos through the group environment remains unchanged. The selective
destruction of halos by tidal stripping has the potential to decrease the
fraction of dSphs at large radii, but as we argue below the dominant uncertainty
at large radii is variation between halo realizations, and thus we do not impose
a more complex tidal destruction criteria. We discuss tidal destruction further
in the conclusions.

\subsection{Merger Trees}
\begin{figure*}[t]
\epsscale{1.0}
\plotone{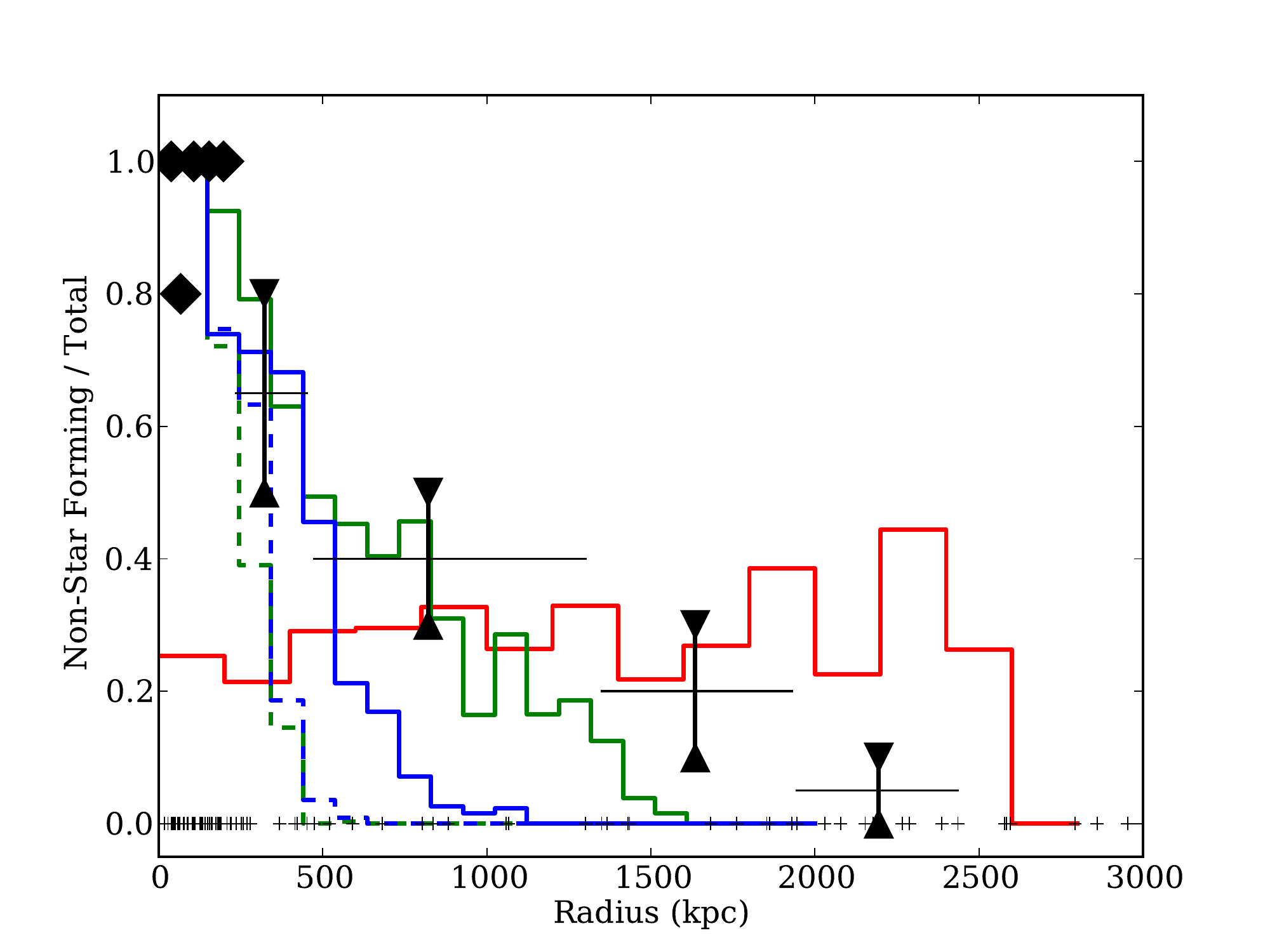}
\caption{Fraction of dSph galaxies as function of galactic radius (minimum of
either Galactocentric or M31-centric), grouped into bins of ten dwarfs and
plotted with black symbols. Upper and lower black triangles differ by including
or excluding intermediate type dwarfs as non-star forming. The span of radius
covered by each bin is shown by the horizontal black lines, and the black points
are plotted at the mean radius of that bin. The green lines show the
fraction of halos that have passed inside $R_{vir}/3$ in VL2, either once (solid
green) or through more than one pericentric passage (dashed green). The blue
lines show the same values but for VL1. The red line shows the distribution of
dwarfs that have undergone major mergers. The horizontal series of ticks along
the bottom indicate the positions of the dwarfs in the sample. The Magellanic
Clouds are included in the bin at 50 kpc.\label{fig_obs}}
\end{figure*}

The fraction of galaxies that have experienced major mergers is calculated from
the same $z=0$ sample and uses the same method of linking halos at each snapshot
to their possible progenitors. However, in the merger trees the selection
requirement for the number of dark matter particles shared between halos is
relaxed, since we are interested in all progenitor halos and not only the most
massive progenitor. Starting at $z=0$, we traverse the merger tree following the
most massive progenitor at each step, until locating a halo that has two
progenitors in a $3:1$ dark matter mass ratio or greater. The timestep where the
two halos are identified as a single 6DFOF halo is noted as the merger time.

Visual inspection of the merger trees suggests that this simple criteria is
effective in identifying whether a halo has merged or not, even though the
merger process may be more complex in detail. Halos can often undergo close
passages, which can cause particles to be lost from the halos by tidal stripping
and thus alter the mass ratio we measure at the final coalescence of the 6DFOF
halos. Other halos undergo passages that temporarily appear as one 6DFOF group in
a snapshot, even though they will later separate and re-coalesce in subsequent
snapshots.  Because we track the time of the most recent merger snapshot, in
these cases our merger times will tend to reflect this final coalescence rather
than initial passes. We are also limited by only tracing the dark matter; we
cannot say when the baryonic components of these galaxies will merge. In general
we expect that when the dark matter halos merge, the baryons must follow, but
this should be delayed by the time required for dynamical friction to bring the
baryonic components together. We present a simple calculation of the dynamical
friction timescale in Section~\ref{conclusions} and find that it is of order 200
Myr or less, which is much smaller than the offset in formation times between
the merger model and the tidal processing model.

Our ability to resolve mergers at very high redshifts is also limited. Beyond
$z>2.5$ (11.3 Gyr ago), halos are poorly linked in time and mergers may not be
properly resolved while they undergo an initial phase of rapid assembly. We
consequently do not track mergers before $z=2.5$. Since there is significant
merger activity near these redshifts, we note that total fraction of dwarfs that
have undergone mergers could be sensitive to the exact cut-off we select, and
consequently we focus primarily on the distribution of merged dwarfs rather than
their absolute fraction. As we will show, the shape of the radial distribution
is unaffected by varying this high redshift cut-off.

\section{Comparison to Observations}
\label{obs}

We compare these simple models to the observed set of Local Group dwarf
galaxies, using the catalog assembled by \citet{mcconnachie12}, which includes
all of the known galaxies within 3 Mpc of the Sun. The catalog labels galaxies
with $M_V > -18$ as dwarfs by convention, and though this cutoff is somewhat
arbitrary, we use the same criterion here. This excludes, for example, M32 and
the Large Magellanic Cloud, but includes the Small Magellanic Cloud. The catalog
provides both galactocentric and M31-centric distances, and also classifies
galaxies as either dSphs, dIrrs, or an intermediate ``dSph/dIrr'' class. Most of
these classifications are uncontroversial. The dSph/dIrr class contains most of
the galaxies for which either observational uncertainty or peculiar combinations
of properties makes it difficult to definitively call them either a dSph or a
dIrr.  Since it is beyond the scope of this work to reconsider the
classification of each of these galaxies, we treat the classification of
\citet{mcconnachie12} as authoritative. We account for the uncertainity in the
dSph/dIrr class by evaluating two scenarios: one where all of these galaxies are
treated as dSphs, and one where they are all treated as dIrrs. The range of
values produced by these two cases yields some estimate of the uncertainty from
classification. We make only two updates to the classifications of
\citet{mcconnachie12} based on more recent works: the galaxy Andromeda XXVIII
has been confirmed to be a dSph (Slater et al., in prep), as has the galaxy KKR
25 \citep{makarov12}. It is important to note that the set of known dwarfs is
not complete, and there may be underlying observational biases in the catalog.
In this work we do not attempt to correct for biases in the selection function.
We make the assumption that dSphs and dIrrs are equally likely to be detected,
and thus the relative fraction of these two types is independent of the
selection function.  Inside of roughly 800 kpc this condition in general is met,
since both dSphs and dIrrs can be detected by their red giant branch stars in
the available large surveys \citep[e.g.,][]{irwin07,slater11}. Outside of this
range dIrrs may be preferentially detected, since their young stars can be
brighter than the tip of the red giant branch in dSphs. We remain mindful of
this potential bias when interpreting the observations, but do not believe it
affects our results.  Our conclusions are necessarily more cautious at very
large radii.

For each dwarf we compute the minimum of either its Galactocentric or its
M31-centric distance, since we are not concerned with which galaxy the dwarfs
may have interacted with. The dwarfs are then binned into groups of ten, and the
fraction of dSphs in each bin is plotted as the black symbols in
Figure~\ref{fig_obs} at the mean radius of its constituent dwarfs. This fixed-number
rather than fixed-width binning scheme is used to compensate for the large
dynamic range in the number of dwarf galaxies as a function of radius. For each
bin, we evaluate the non-star forming fraction with the intermediate dSph/dIrr
type galaxies included as dSphs (upper triangles) and as dIrrs (lower
triangles), and the two points are connected by the vertical black lines. Bins
with no transition galaxies appear as diamonds. The range of radius values
spanned by each grouping of ten dwarfs is shown by the black horizontal lines on
each point.

In Figure~\ref{fig_obs}, the red line shows the radial distribution of Via
Lactea II halos that have had a major merger since $z=2.5$. This distribution is
clearly flat, and does not exhibit the rise in non-star forming dwarfs inside of
1 Mpc as is seen in the Local Group. \textit{This radial dependence alone
suggests that mergers cannot be the only channel for dSph formation}.

Also in Figure~\ref{fig_obs}, the fraction of satellite halos from the
simulations that have passed inside of $R_{vir}/3$ is shown by the solid blue
line for VL1 and the green line for VL2. (For VL2 this also uses the minimum
distance between a halo at $z=0$ and either the main host halo or Halo 2.) This
is a simple proxy for the dwarfs that could have undergone transformation by a
tidal interaction.

The radial profile in VL2 agrees quite well with the distribution of observed
dSphs, with a gradual decline in non-star-forming fraction from 400 to 1500
kpc. This illustrates that tidal processes can plausibly reproduce the observed set
of dSphs. However, the VL1 profile falls off much more rapidly, with very few
tidally processed halos found beyond 800 kpc. The difference between the VL1 and
VL2 results suggests that the predicted radial profile of dSphs in this model is
clearly not a smooth, universal function. There is a large stochastic component
that is evident even with only two realizations of a Local Group-like
environment, which produces variations in the dSph profile beyond what would be
expected from just Poisson noise. This variation comes from the accretion of
subgroups of halos, which follow similar trajectories and introduce correlations
in the fraction of processed halos. The accretion of discrete subgroups has been
seen in other simulations, such as \citet{li08} and \citet{klimentowski10}, and
we include a more detailed illustration of this effect in
Section~\ref{subgroups}. With only two realizations we are unable to quantify
this effect beyond showing the two simulations as illustrating the possible
magnitude of variations. 

The confirmed dSph KKR 25 at 1.9 Mpc is the most significant outlier from the
agreement between the observations and the simulations. Though this discrepancy
could result from our two simulations failing to span the entire range of
possible outcomes, it is also possible that our simple criteria for forming
dSphs is imprecise and a more lenient criteria could account for KKR 25. With
these caveats it is difficult to convincingly argue that tidal processing cannot
account for KKR 25, but it is an interesting test case that could be suggestive
of merger activity. As discussed above, the normalization on the fraction of
merged dwarfs is somewhat sensitive to the details of the merger criteria,
primarily the upper redshift cutoff and the mass ratio of merger required. While
this sensitivity and the limitations of Poisson noise limit our ability to draw
conclusions about whether the two furthest bins are compatible with any
merger-based dSph formation, the figure does show the range of radii over which
the two formation scenarios could be active.

The dashed blue and green lines in Figure~\ref{fig_obs} take the same tidal
processing criteria as the solid lines, but adds an additional constraint that
the halo must experienced more than one pericentric passage inside of
$R_{vir}/2$. This is slightly less restrictive in distance than the single pass
criterion, since multiple weaker tidal interactions could replace a single strong
interaction. As shown in the figure, the fraction of halos in either simulation
with two or more passages drops steeply outside of 300 kpc, and is essentially
zero beyond 500 kpc. This agrees with the results of a simple orbital timescale
calculation at these radii, which shows that the single orbits require a
significant fraction of a Hubble time.  Performing this test in a cosmological
simulation accounts for more complicated factors such as the growth the main
halo and the initial positions and velocities of the halos that are today found
at these radii. The result of the simulation clearly shows that dwarfs such as
Cetus, Tucana, and KKR 25 could not have made multiple close passages by a large
galaxy; if they were transformed into dSphs by tidal forces, it must have been
done by a single passage.

\begin{figure*}
\plotone{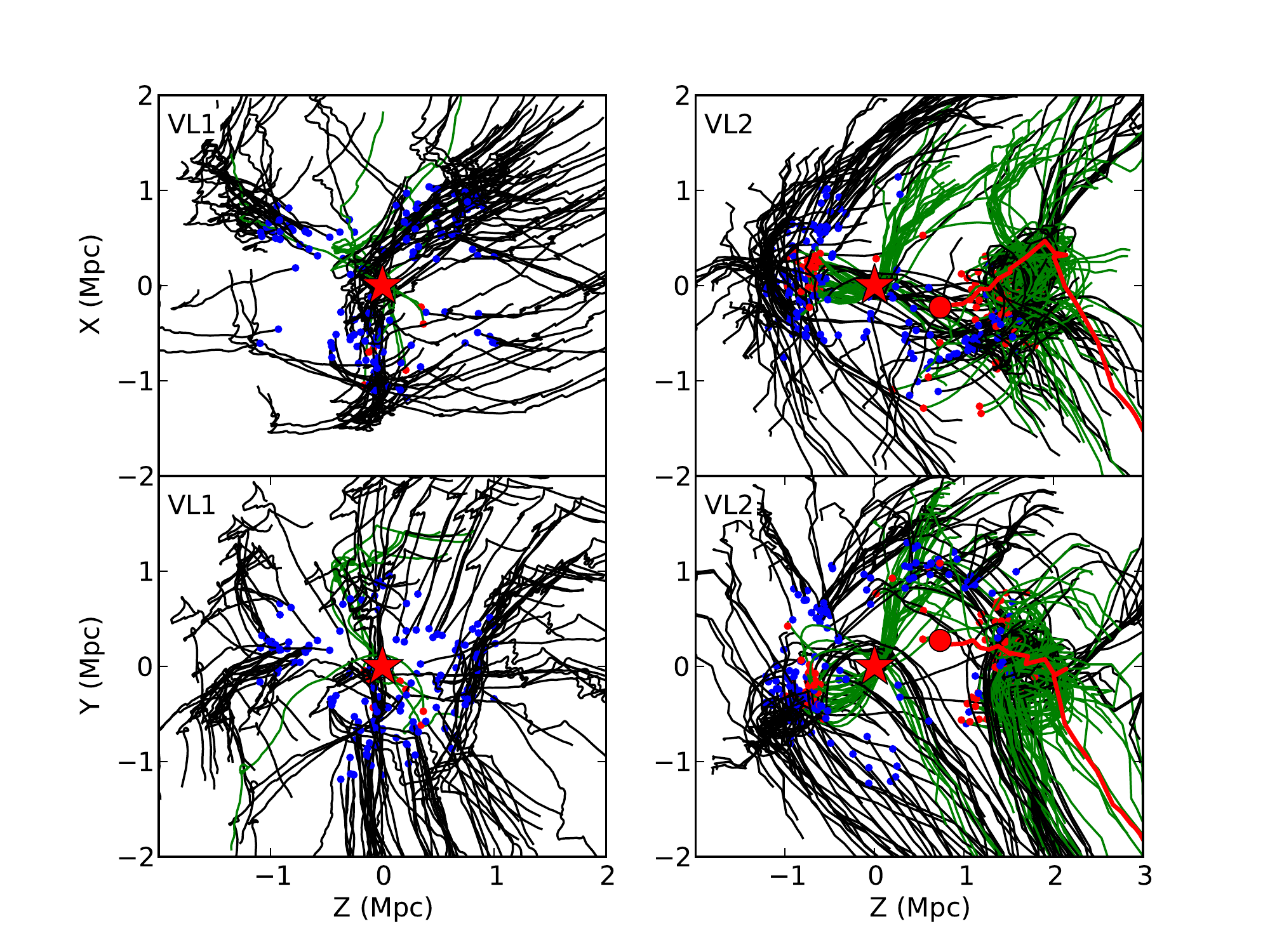}
\caption{Trajectory of present-day distant halos in VL1 (left) and VL2 (right),
in comoving coordinates relative to the motion of the main halo which is held
fixed at the origin. The $z=0$ location of the halos are marked with dots. Halos
that have passed inside $R_{vir}/3$ are marked with green lines and a red dot,
while all others are marked with black lines and a blue dot. Halo 2 in VL2 is
marked by the large red dot and the red track. The clumpy nature of the
accretion is clearly visible. VL1 has several subgroups which have not passed by
the main halo yet, while several of the subgroups in VL2 clearly have and are
receding from the main halo.\label{tracks}}
\end{figure*}

\subsection{Accretion History}
\label{subgroups}

In discussing Figure~\ref{fig_obs} it was argued that the significant difference in
the histories of halos in VL1 and VL2 was due to coherent subgroups of halos.
In VL2 several of these subgroups had passed near the host galaxy, while in VL1
very few did. This difference is illustrated in Figure~\ref{tracks}, which shows
the trajectories in comoving coordinates of halos that have $z=0$ radii of
500-1500 kpc. The trajectories are all relative to the host galaxy, which is
fixed at the origin (denoted by the red star). The trajectories of halos that
have not passed inside $R_{vir}/3$ are shown in black with blue dots at their
$z=0$ position, while those that have are shown with green lines and red dots.
In the VL2 panels, the trajectory of Halo 2 is shown in red with a large red
dot.

In both simulations it is clear that many halos are organized into small groups
with correlated trajectories. The left side of the VL1 plots show one of these
groupings clearly. The same effect is shown in VL2, most clearly seen in the
paths, but with the distinct difference that several of these groups have passed
through the main halo (or Halo 2), and have thus potentially been tidally
processed. This correlated nature of the infalling halos is what causes the
significant variation in the radial profile of processed halos, above and beyond
what would be expected from pure Poisson noise on the individual halos. Infall
of small subgroups of dwarfs has been seen in many other simulations
\citep{li08,klimentowski10,lovell11,helmi11} and has been argued to be the cause
of the apparent position or velocity correlations amongst satellites around the
Milky Way \citep{lyndenbell76,libeskind05,fattahi13}, M31
\citep{ibata13,conn13}, and more distant neighbors of the Local Group
\citep{tully06}.

\subsection{Parameter Sensitivity}

\begin{figure}
\epsscale{1.3}
\plotone{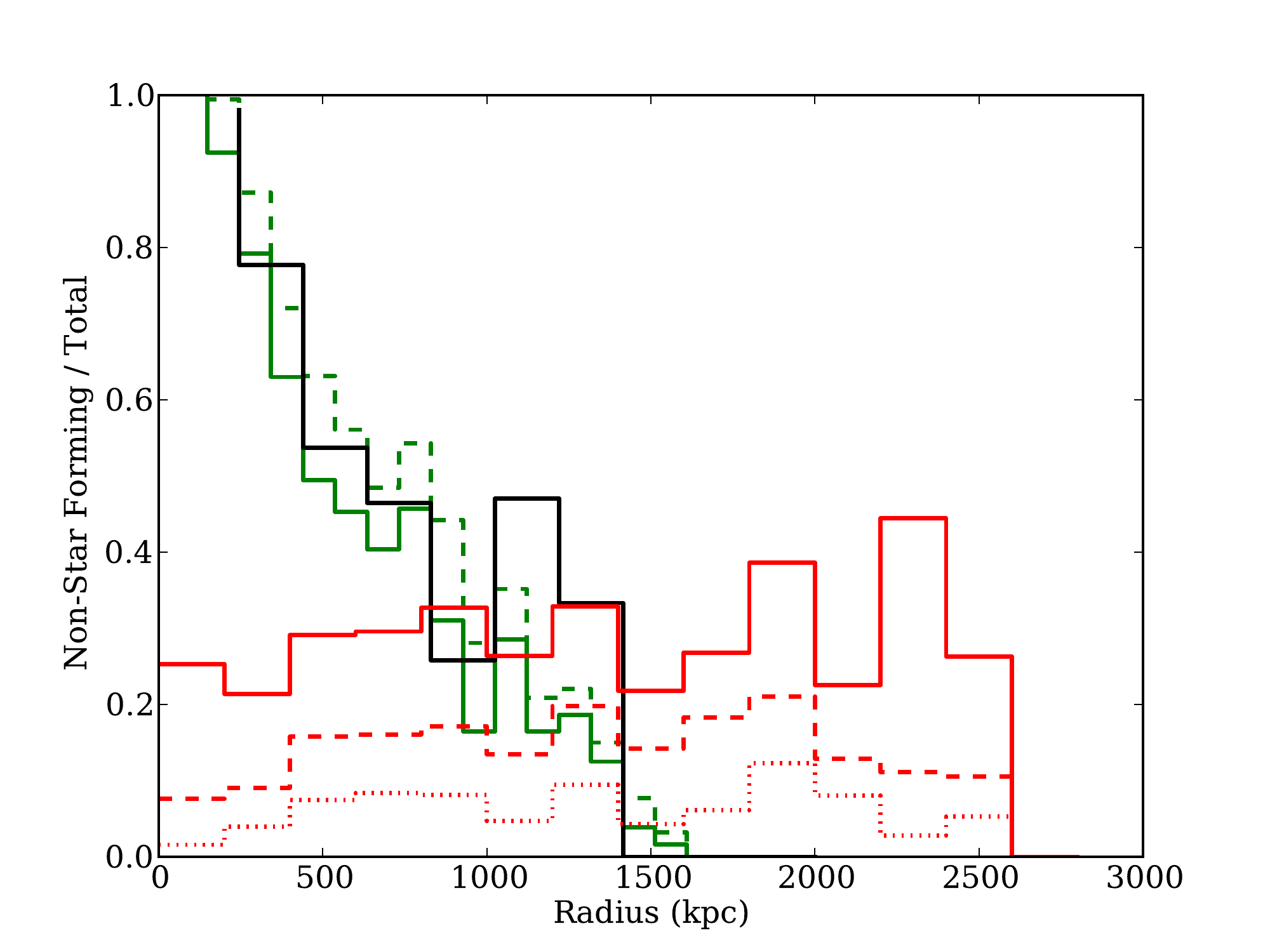}
\caption{Illustration of the sensitivity of our results to to various parameter
choice, using the VL2 data. The solid green line is the same as in
Figure~\ref{fig_obs}, while the dashed green line shows those that have passed
inside $R_{vir}/2$ instead of $R_{vir}/3$. The black line illustrates changing
the \vmax criterion to $V_{\rm max} > 10$ km/s rather than 5 km/s. The solid,
dashed, and dotted red lines show halos that have undergone mergers in the last
11.3, 10, and 8 Gyr, respectively.  All of these variations may change the
normalization of the model results, but do not affect the general form.
\label{variation}}
\end{figure}

Though our analysis includes some fixed parameters that could potentially alter
the results, the robustness of the general conclusions can be shown by
recalculating the results under slightly different assumptions.
Figure~\ref{variation} shows the result of changing these assumptions. The solid
green line is the same as used in Figure~\ref{fig_obs}, while the dashed green line
shows the same calculation but under the relaxed assumption that a galaxy could
be tidally affected inside $R_{vir}/2$, rather than $R_{vir}/3$. This increases
the non-star forming fraction at all radii (as it must), but shows a
similarly-shaped radial dependence. The solid black line in
Figure~\ref{variation} also shows the same calculation as before, but tightening
the \vmax constraint to only include halos with $V_{max} > 10$ km/s rather than
$5$ km/s. This includes many fewer halos, so the resulting plot is more noisy
and we have had to double the bin size accordingly, but again the radial
dependence is similar. 

The most significant parameters in the merger calculation are the redshift
cut-off and the required merger mass ratio. Both of these alter the
absolute number of dwarfs that have undergone mergers without altering the
$z=0$ radial distribution. This is shown by the red lines in
Figure~\ref{variation}, where the solid line shows dwarfs with mergers more
recent than 11 Gyr, the dashed shows those more recent than 10 Gyr, and the
dotted corresponds to 8 Gyr ago. The number of dwarfs with mergers drops by over
half in the most restrictive of these cases, but no other effects are seen. Our
analysis remains cognizant of this effect and thus it should not compromise our
conclusions.

\section{Transformation Timescales}
\label{times}

\begin{figure}
\epsscale{1.3}
\plotone{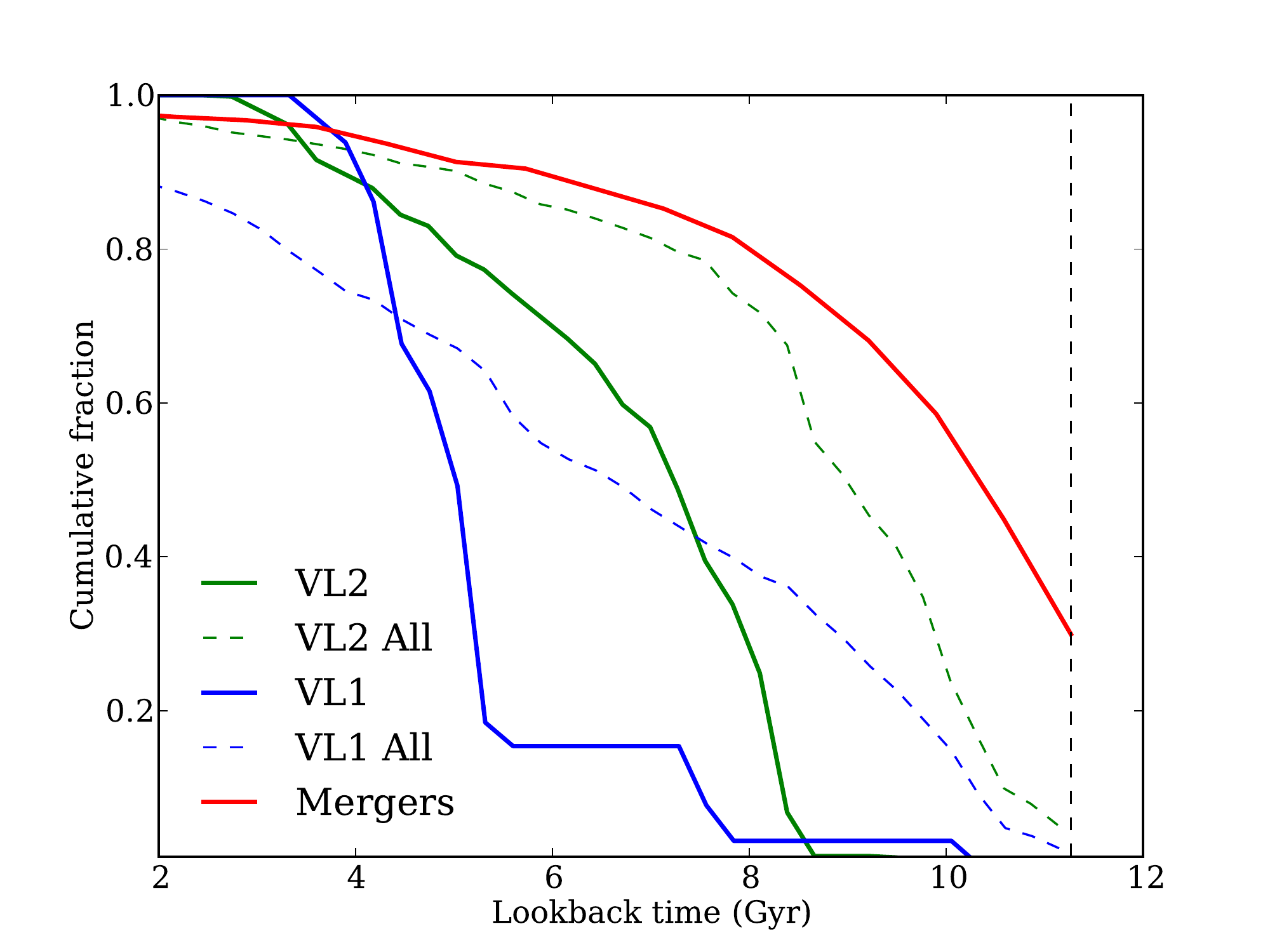}
\caption{Cumulative distribution of the timescales at which the selected $z=0$
halos first met various dSph formation criteria. The blue (VL1) and green (VL2)
solid lines show when halos the halos today found between 500 and 1500 kpc from
their host first crossed $R_{vir}/3$. The dashed lines show the same
calculation for each simulation, but without the present-day radius restriction.
The red line shows the time at which dark matter halos (those at $z=0$ between
500 and 1500 kpc) first underwent a major merger. The vertical dashed line
indicates the time at which we are first able to resolve mergers or close
passages. \label{timescales}}
\end{figure}

By tracing the merger and the tidal transformation scenarios with cosmological
simulations, we are able to infer the timescales on which either of these
processes would have been active. The time at which star formation stopped in
the dwarf is imprinted in the stellar populations and could be used to
differentiate between the two scenarios for dSph formation.
Figure~\ref{timescales} shows the cumulative distribution of times at which the
distant halos of our $z=0$ sample (located between 500 and 1500 kpc from either
host galaxy) either underwent its most recent merger (red line) or first met the
tidal criteria (passing inside $R_{vir}/3$, shown as the solid blue line for VL1
and sold green for VL2). The vertical dashed line indicates the first timestep
at which we are able to resolve mergers or close passages. For comparison we
also show the distribution of times at which surviving halos at any present day
radius first crossed $R_{vir}/3$ (dashed blue VL1, dashed green VL2).

The distribution of merger times is clearly weighted towards early times. From
the plot, roughly 50\% of the observed dwarf-sized dark matter halos that
experienced mergers did so more than 11 Gyr ago. This is partly due to the epoch
of assembly for small halos being biased towards early times, but there is also
the factor of the small halos' infall onto the larger host increasing the
relative velocities of halos to each other and thus inhibiting dynamical
friction and further merging.

Though we can directly measure the time at which dark matter halos merged in the
simulation, we must also account for the fact that the baryonic components of
these galaxies may require additional time for dynamical friction to bring the
baryons to coalescence. This is not directly observable, but we can estimate the
time lag between the merger of the dark matter and the baryons with the
dynamical friction formula from \citet{BT08},
\begin{equation}
t_{\rm fric} = \frac{2.34}{\ln \Lambda} \left(\frac{\sigma_h}{\sigma_s}\right)^2
\frac{r}{\sigma_s},
\end{equation}
where $\sigma_h$ and $\sigma_s$ are the velocity dispersions of the ``host'' and
``satellite'' halos, $r$ is a characteristic radius over which dynamical
friction must act to bring the baryonic components together, and $\ln \Lambda$
is the Coulomb logarithm. Our selection of only major mergers constrains
$\sigma_h/\sigma_s$ to be roughly the square root of the mass ratio, and the
Coulomb logarithm can be calculated as $\Lambda = 2^{3/2} \sigma_h/\sigma_m$
\citep{BT08}. The dynamical friction time thus reduces to a small factor of
crossing time. An order of magnitude estimate with $r \sim 1$ kpc and $\sigma
\sim 10$ km/s yields $t_{\rm fric} = 200$ Myr. Assuming a delay of 1 Gyr between
the dark matter merger and the baryonic merger would be a relatively
conservative estimate.

Given that we see significant merger and accretion activity occurring at such
early times, it is logical to ask why the infall of distant dwarfs ($z=0$
radii of 500-1500 kpc) onto the host galaxy is so delayed. For example, the
rapid rise of the VL2 cumulative infall fraction in Figure~\ref{timescales}
seems to start suddenly between 8--9 Gyr ago. Two points can help explain this.
First, as discussed above, the accretion of halos onto the host galaxy is
stochastic and several ``clumps'' of subhalos are sometimes accreted together.
This effect contributes to the stochasticity of the infall rate, particularly in
the VL1 simulation where there are fewer halos found at large radii at $z=0$.
The other effect that causes the delayed infall times is related to the distance
cut we have applied. The solid green and blue lines in Figure~\ref{timescales}
only shows halos that today are found between 500 and 1500 kpc; for comparison,
the dashed lines shows the same infall calculation but with that present-day
radius constraint removed. In the VL2 case, by the time the infall of the
present-day distant dwarfs is starting more than 70\% of all surviving halos
have already been accreted onto the host galaxy. These halos accreted at early
times join a galaxy which is much smaller at the time of accretion, and thus fall
in close to the galaxy, while satellites that fall in at later times encounter a
much larger galaxy which has grown around the close-in satellites. The VL1 case
is more stochastic and the accretion is weighted towards even later times, but
the delay is still present. Though this is merely a rough sketch to illustrate
the process, the simulation is clear in predicting late infall times for distant
dwarfs. When contrasted with the timescale for the merger scenario, the
simulations clearly point to a difference in formation times for the two
channels.

\section{Discussion and Conclusions}
\label{conclusions}

We have shown that the radial distribution of galaxies of dwarf galaxies of
different morphological types can be used to constrain their formation
mechanisms, when combined with simple models based on cosmological simulations.
The simulations show that these models produce substantially different sets of
properties for the Local Group dwarfs. Mergers of dwarfs are clearly
insufficient to explain all of the dSphs, and tidal processes that require
multiple pericenter passages cannot account for the number of dwarfs found
further than 500 kpc from their host galaxy. These two points are robust.

The fact that the simple close passage model is able to reproduce the observed
radial profile of dSphs, even if it does not do so in all cosmological
realizations, suggests that this model could be sufficient to create the
observed dSphs. However, it hinges critically on the single-passage
requirement. Many simulations have shown that multiple passages are necessary
with some dwarf models \citep{mayer01,kazantzidis11a}, though the more recent
simulations of \citet{kazantzidis13} suggest that a single-passage
transformation is plausible if the progenitor dwarfs have shallow mass profiles
in the center (cores). Such profiles contradict early predictions of Cold Dark
Matter models \citep{dubinski91,navarro96,navarro97}, but observational studies
have shown that cores are prevalent in dwarf galaxies \citep{walker11,oh11} and
simulations \citep{read05,governato10,governato12,zolotov12} have shown
reasonable methods to create cores from baryonic processes. Cored profiles,
however, carry the risk that the halos are more susceptible to tidal stripping
and even complete destruction \citep{penarrubia10}. The fine balance between
tidal transformation/stirring and tidal destruction may further constrain
distant dSphs to a narrow range of structural and orbital parameters.

The alternative formation pathway we have studied, that of mergers between
dwarfs, has less evidence to support it but is difficult to rule out. We show
that it is unable to be the dominant pathway by which dSphs form, simply because
it does not recreate the large dSph fraction at small radii to a host galaxy.
However, assuming our understanding of dark matter is correct, mergers must
occur. Whether these mergers leave signatures that are observable today is a
challenging question that requires further study. At very high redshifts dwarfs
may be able to reform gas disks and continue forming stars, which would lead us
to identify them as dIrrs. At what redshift, if any, mergers cause these
galaxies to no longer sustain any star formation is a complex question best
answered with hydrodynamical simulations.

We have shown that, for distant dwarfs, mergers must occur at very early times,
while their infall onto a host potential occurs much later. The time at which
star formation ended in the dwarf should therefore be a signature in the stellar
populations that cannot be erased. Do the distant dSphs of the Local Group
exhibit star formation histories that could differentiate between the early
shut-off in the merger scenario and the later shut-off by tidal interactions?

Studies of the star formation histories of Tucana and Cetus show that both
dwarfs reached the peak star formation rate more than $12$ Gyr ago and
subsequently the star formation rates declined \citep{monelli10a,monelli10b},
but from this history it is difficult to ascribe a specific time at which some
process shut off star formation. In either case it took nearly 3 Gyr for the
star formation rate to decline from its peak value to negligible levels;
such a slow and gradual process does not lend itself to an easy comparison to
our binary off-or-on model. This is particularly true in the case of mergers,
where additional star formation may be triggered by the merger itself. More
sophisticated modeling of the detailed star formation history, including the
hydrodynamical processes that eventually render a dSph devoid of gas, could be
able to extract conclusions from the stellar populations seen in Cetus and
Tucana.

A logical extension of our work would be to ask if there exist comparable trends
outside the Local Group. The work of \citet{geha12} used a sample of somewhat
more massive dwarfs to show that below a mass threshold of $10^9$ $M_\odot$,
non-star-forming galaxies do not exist in any substantial number beyond 1500 kpc
of a massive galaxy. This matches well with the predictions of the tidal
processing scenario, which also shows very few processed halos beyond 1500
kpc. The \citet{geha12} sample substantiates the hypothesis that at low masses
tidal processing is sufficient to recreate the distribution of dSphs, without
requiring mergers. At slightly higher masses of $10^{9.5}-10^{9.75} M_\odot$, a
small fraction of quenched halos are observed at all radii (their Figure~4),
much in agreement with the expectations for mergers. If mergers are responsible
at large radii, this suggests that it is only above a certain mass threshold
that mergers (which must happen at all masses) are capable of quenching galaxies
in the field. Such a model has been shown by \citet{hopkins09} to reproduce the
mass dependence of the fraction of bulge-dominated galaxies in the field, and
could extend to their star formation properties as well. This picture of
combining tidal processes with mergers above a threshold provides a natural link
between the behavior of satellites and of central galaxies. 

One difference between the \citet{geha12} results and the Local Group is in the
fraction of quenched galaxies at small radii. In the Local Group 
substantially all galaxies inside 200 kpc are quenched, but the quenched
fraction only reaches at most 30\% in the \citet{geha12} sample. This could
suggest a mass dependence to tidal processing, where perhaps the more massive
dwarfs of the \citet{geha12} sample require longer timescales to shut off star
formation, and thus many of their galaxies at radii are slowly on the way to
quenching. Our instantaneous tidal processing model does not capture this
behavior, but a more sophisticated mass-dependent model may better explain this
effect.

\acknowledgments

This work was supported by NSF grant AST 1008342. We thank the Via Lactea
collaboration for making their simulation outputs available, and J. Diemand for
providing the merger trees and helpful comments on their interpretation. We also
thank J. Dalcanton for useful discussions.


\begin{thebibliography}{}
\bibitem[Binney \& Tremaine(2008)]{BT08} Binney, J., \& Tremaine, S.\ 2008,
Galactic Dynamics: Second Edition, Princeton University Press, Princeton, NJ
\bibitem[Balogh et al.(2000)]{balogh00} Balogh, M.~L., Navarro, J.~F., \&
Morris, S.~L.\ 2000, \apj, 540, 113 
\bibitem[Brooks et al.(2013)]{brooks13} Brooks, A.~M., Kuhlen, M., Zolotov, A.,
\& Hooper, D.\ 2013, \apj, 765, 22 
\bibitem[Conn et al.(2013)]{conn13} Conn, A.~R., Lewis, G.~F., Ibata, R.~A., et
al.\ 2013, \apj, 766, 120 
\bibitem[Dekel \& Silk(1986)]{dekel86} Dekel, A., \& Silk, J.\ 1986, \apj, 303, 39 
\bibitem[Diemand et al.(2006)]{diemand06} Diemand, J., Kuhlen, M., \& Madau, P.\
2006, \apj, 649, 1 
\bibitem[Diemand et al.(2007)]{diemand07} Diemand, J., Kuhlen, M., \& Madau,
P.\ 2007, \apj, 667, 859
\bibitem[Diemand et al.(2008)]{diemand08} Diemand, J., Kuhlen, M., Madau, P., et
al.\ 2008, \nat, 454, 735 
\bibitem[D'Onghia et al.(2009)]{donghia09} D'Onghia, E., Besla, G., Cox, T.~J.,
\& Hernquist, L.\ 2009, \nat, 460, 605 
\bibitem[Dubinski \& Carlberg(1991)]{dubinski91} Dubinski, J., \& Carlberg,
R.~G.\ 1991, \apj, 378, 496
\bibitem[Faber \& Lin(1983)]{faber83} Faber, S.~M., \& Lin, D.~N.~C.\ 1983,
\apjl, 266, L17 
\bibitem[Fattahi et al.(2013)]{fattahi13} Fattahi, A., Navarro, J.~F.,
Starkenburg, E., Barber, C.~R., \& McConnachie, A.~W.\ 2013, \mnras, 431, L73 
\bibitem[Gallagher \& Wyse(1994)]{gallagher94} Gallagher, J.~S., III, \& Wyse,
R.~F.~G.\ 1994, \pasp, 106, 1225 
\bibitem[Geha et al.(2012)]{geha12} Geha, M., Blanton, M.~R., Yan, R., \&
Tinker, J.~L.\ 2012, \apj, 757, 85 
\bibitem[Gill et al.(2005)]{gill05} Gill, S.~P.~D., Knebe, A., \& Gibson, B.~K.\
2005, \mnras, 356, 1327 
\bibitem[Gnedin(2000)]{gnedin00} Gnedin, N.~Y.\ 2000, \apj, 542, 535 
\bibitem[Gnedin \& Zhao(2002)]{gnedin02} Gnedin, O.~Y., \& Zhao, H.\ 2002,
\mnras, 333, 299 
\bibitem[Governato et al.(2010)]{governato10} Governato, F., Brook, C., Mayer,
L., et al.\ 2010, \nat, 463, 203 
\bibitem[Governato et al.(2012)]{governato12} Governato, F., Zolotov, A.,
Pontzen, A., et al.\ 2012, \mnras, 422, 1231 
\bibitem[Grebel et al.(2003)]{grebel03} Grebel, E.~K., 
Gallagher, J.~S., III, \& Harbeck, D.\ 2003, \aj, 125, 1926 
\bibitem[Helmi et al.(2011)]{helmi11} Helmi, A., Cooper, A.~P., 
White, S.~D.~M., et al.\ 2011, \apjl, 733, L7 
\bibitem[Hodge(1964)]{hodge64} Hodge, P.~W.\ 1964, \aj, 69, 438
\bibitem[Hodge \& Michie(1969)]{hodge69} Hodge, P.~W., \& Michie, R.~W.\ 1969,
\aj, 74, 587 
\bibitem[Hopkins et al.(2009)]{hopkins09} Hopkins, P.~F., Somerville, R.~S.,
Cox, T.~J., et al.\ 2009, \mnras, 397, 802 
\bibitem[Ibata et al.(2013)]{ibata13} Ibata, R.~A., Lewis, G.~F., Conn, A.~R.,
et al.\ 2013, \nat, 493, 62 
\bibitem[Irwin et al.(2007)]{irwin07} Irwin, M.~J., Belokurov, V., Evans, N.~W.,
et al.\ 2007, \apjl, 656, L13 
\bibitem[Kazantzidis et al.(2011a)]{kazantzidis11a} Kazantzidis, S., {\L}okas,
E.~L., Callegari, S., Mayer, L., \& Moustakas, L.~A.\ 2011, \apj, 726, 98 
\bibitem[Kazantzidis et al.(2011b)]{kazantzidis11b} Kazantzidis, S., {\L}okas,
E.~L., Mayer, L., Knebe, A., \& Klimentowski, J.\ 2011, \apjl, 740, L24 
\bibitem[Kazantzidis et al.(2013)]{kazantzidis13} Kazantzidis, S., 
{\L}okas, E.~L., \& Mayer, L.\ 2013, \apjl, 764, L29 
\bibitem[Klimentowski et al.(2009)]{klimentowski09} Klimentowski, J., {\L}okas,
E.~L., Kazantzidis, S., Mayer, L., \& Mamon, G.~A.\ 2009, \mnras, 397, 2015 
\bibitem[Klimentowski et al.(2010)]{klimentowski10} Klimentowski, J., 
{\L}okas, E.~L., Knebe, A., et al.\ 2010, \mnras, 402, 1899 
\bibitem[Kormendy(1985)]{kormendy85} Kormendy, J.\ 1985, \apj, 295, 73 
\bibitem[Klypin et al.(1999)]{klypin99} Klypin, A., Kravtsov, A.~V., Valenzuela,
O., \& Prada, F.\ 1999, \apj, 522, 82 
\bibitem[Koposov et al.(2009)]{koposov09} Koposov, S.~E., Yoo, J., Rix, H.-W.,
et al.\ 2009, \apj, 696, 2179 
\bibitem[Kravtsov et al.(2004)]{kravtsov04} Kravtsov, A.~V., Gnedin, O.~Y., \&
Klypin, A.~A.\ 2004, \apj, 609, 482 
\bibitem[Li \& Helmi(2008)]{li08} Li, Y.-S., \& Helmi, A.\ 2008, \mnras, 385,
1365 
\bibitem[Libeskind et al.(2005)]{libeskind05} Libeskind, N.~I., Frenk, C.~S.,
Cole, S., et al.\ 2005, \mnras, 363, 146 
\bibitem[Lovell et al.(2011)]{lovell11} Lovell, M.~R., Eke, 
V.~R., Frenk, C.~S., \& Jenkins, A.\ 2011, \mnras, 413, 3013 
\bibitem[Lynden-Bell(1976)]{lyndenbell76} Lynden-Bell, D.\ 1976, \mnras, 174,
695 
\bibitem[Mac Low \& Ferrara(1999)]{maclow99} Mac Low, M.-M., \& Ferrara, A.\
1999, \apj, 513, 142 
\bibitem[Makarov et al.(2012)]{makarov12} Makarov, D., Makarova, L., Sharina,
M., et al.\ 2012, \mnras, 425, 709 
\bibitem[Mayer et al.(2001)]{mayer01} Mayer, L., Governato, F., 
Colpi, M., et al.\ 2001, \apj, 559, 754 
\bibitem[Mayer et al.(2006)]{mayer06} Mayer, L., Mastropietro, 
C., Wadsley, J., Stadel, J., \& Moore, B.\ 2006, \mnras, 369, 1021 
\bibitem[McConnachie(2012)]{mcconnachie12} McConnachie, A.~W.\ 2012, \aj, 144, 4 
\bibitem[Monelli et al.(2010a)]{monelli10a} Monelli, M., Hidalgo, 
S.~L., Stetson, P.~B., et al.\ 2010, \apj, 720, 1225 
\bibitem[Monelli et al.(2010b)]{monelli10b} Monelli, M., Gallart, 
C., Hidalgo, S.~L., et al.\ 2010, \apj, 722, 1864 
\bibitem[Moore(1994)]{moore94} Moore, B.\ 1994, \nat, 370, 629 
\bibitem[Moore et al.(1999)]{moore99} Moore, B., Ghigna, S., Governato, F., et
al.\ 1999, \apjl, 524, L19 
\bibitem[Moore et al.(2004)]{moore04} Moore, B., Diemand, J., \& Stadel, J.\
2004, IAU Colloq.~195: Outskirts of Galaxy Clusters: Intense Life in the
Suburbs, 513 
\bibitem[Navarro et al.(1996)]{navarro96} Navarro, J.~F., Frenk, 
C.~S., \& White, S.~D.~M.\ 1996, \apj, 462, 563 
\bibitem[Navarro et al.(1997)]{navarro97} Navarro, J.~F., Frenk, C.~S., \&
White, S.~D.~M.\ 1997, \apj, 490, 493 
\bibitem[Oh et al.(2011)]{oh11} Oh, S.-H., de Blok, W.~J.~G., Brinks, E.,
Walter, F., \& Kennicutt, R.~C., Jr.\ 2011, \aj, 141, 193 
\bibitem[Pe{\~n}arrubia et al.(2010)]{penarrubia10} Pe{\~n}arrubia, J., Benson,
A.~J., Walker, M.~G., et al.\ 2010, \mnras, 406, 1290 
\bibitem[Read \& Gilmore(2005)]{read05} Read, J.~I., \& Gilmore, G.\ 2005,
\mnras, 356, 107 
\bibitem[Sawala et al.(2010)]{sawala10} Sawala, T., Scannapieco, C., Maio, U.,
\& White, S.\ 2010, \mnras, 402, 1599 
\bibitem[Slater et al.(2011)]{slater11} Slater, C.~T., Bell, E.~F., \& Martin,
N.~F.\ 2011, \apjl, 742, L14 
\bibitem[Somerville(2002)]{somerville02} Somerville, R.~S.\ 2002, \apjl, 572,
L23 
\bibitem[Teyssier et al.(2012)]{teyssier12} Teyssier, M., Johnston, K.~V., \&
Kuhlen, M.\ 2012, \mnras, 426, 1808 
\bibitem[Tully et al.(2006)]{tully06} Tully, R.~B., Rizzi, L., Dolphin, A.~E.,
et al.\ 2006, \aj, 132, 729 
\bibitem[van den Bergh(1994)]{vandenbergh94} van den Bergh, S.\ 1994, 
\apj, 428, 617 
\bibitem[Walker \& Pe{\~n}arrubia(2011)]{walker11} Walker, M.~G., \&
Pe{\~n}arrubia, J.\ 2011, \apj, 742, 20 
\bibitem[Wetzel et al.(2013)]{wetzel13} Wetzel, A.~R., Tinker, J.~L., Conroy,
C., \& van den Bosch, F.~C.\ 2013, arXiv:1303.7231 
\bibitem[Zolotov et al.(2012)]{zolotov12} Zolotov, A., Brooks, A.~M., Willman,
b., et al.\ 2012, \apj, 761, 71 
\end{thebibliography}
\end{document}